\documentclass[aps,onecolumn,amsmath,amssymb,superscriptaddress,PRL]{revtex4-2}

\usepackage{graphicx}
\usepackage{bm}
\usepackage{color}
\usepackage{easyReview}
\usepackage{latexsym} 
\usepackage{natbib}

\usepackage[unicode=true,colorlinks=true]{hyperref}
\DeclareUnicodeCharacter{FF0C}{,}

\hypersetup{linkcolor=blue,citecolor=blue,urlcolor=blue}

\begin{document}

\title{Laughlin pumping assisted by surface acoustic waves}

\author{Renfei Wang}
\affiliation{International Center for Quantum Materials, Peking University, Haidian, Beijing 100871, China}
\author{Xiao Liu}
\affiliation{International Center for Quantum Materials, Peking University, Haidian, Beijing 100871, China}
\author{Adbhut Gupta} 
\affiliation{Department of Electrical Engineering, Princeton University, Princeton, New Jersey 08544, USA}
\author{Kirk W. Baldwin} 
\affiliation{Department of Electrical Engineering, Princeton University, Princeton, New Jersey 08544, USA}
\author{Loren Pfeiffer} 
\affiliation{Department of Electrical Engineering, Princeton University, Princeton, New Jersey 08544, USA}
\author{Wenfeng Zhang}
\affiliation{International Center for Quantum Materials, Peking University, Haidian, Beijing 100871, China}
\author{Rui-Rui Du} 
\affiliation{International Center for Quantum Materials, Peking University, Haidian, Beijing 100871, China}
\affiliation{Hefei National Laboratory, Hefei 230088, China}
\author{Mansour Shayegan} 
\affiliation{Department of Electrical Engineering, Princeton University, Princeton, New Jersey 08544, USA}
\author{Xi Lin} 
\affiliation{International Center for Quantum Materials, Peking University, Haidian, Beijing 100871, China}
\affiliation{Hefei National Laboratory, Hefei 230088, China}
\author{Ying-Hai Wu} 
\affiliation{School of Physics and Wuhan National High Magnetic Field Center, Huazhong University of Science and Technology, Wuhan 430074, China}
\author{Yang Liu} 
\affiliation{International Center for Quantum Materials, Peking University, Haidian, Beijing 100871, China}
\affiliation{Hefei National Laboratory, Hefei 230088, China}


\begin{abstract}
\textbf{The quantum Hall effect is a fascinating electrical transport phenomenon signified by precise quantization of Hall conductivity $\sigma_\mathrm{xy}$ and vanishing longitudinal conductivity $\sigma_\mathrm{xx}$. Laughlin proposed an elegant explanation in which adiabatic insertion of a flux tube pumps charge through the system. This analysis unveils the fundamental role of gauge invariance and provides a compelling argument about the fractional charge of fractional quantum Hall states. While it has been used extensively as a theoretical tool, a quantitative experimental investigation is lacking despite multiple attempts. Here we report successful realizations of Laughlin pumping in several integer and fractional quantum Hall states. One essential technical innovation is using surface acoustic waves to periodically clear the charges accumulated during the pumping process. Magnetic fluxes are inserted at a constant rate so there is no need to perform complicated data fitting. Furthermore, our setting can reliably extract $\sigma_\mathrm{xx}$ that is several orders of magnitude lower than the limit of conventional techniques. Effective energy gaps can be deduced from the temperature dependence of $\sigma_\mathrm{xx}$, which are drastically different from those provided by conventional transport data. This work not only brings a famous gedanken experiment to reality but also serves as a portal for many future investigations.}
\end{abstract}

\maketitle

\noindent\textbf{Introduction} 
\vspace{1em}

\begin{figure*}[!htbp]
\includegraphics[width=\textwidth]{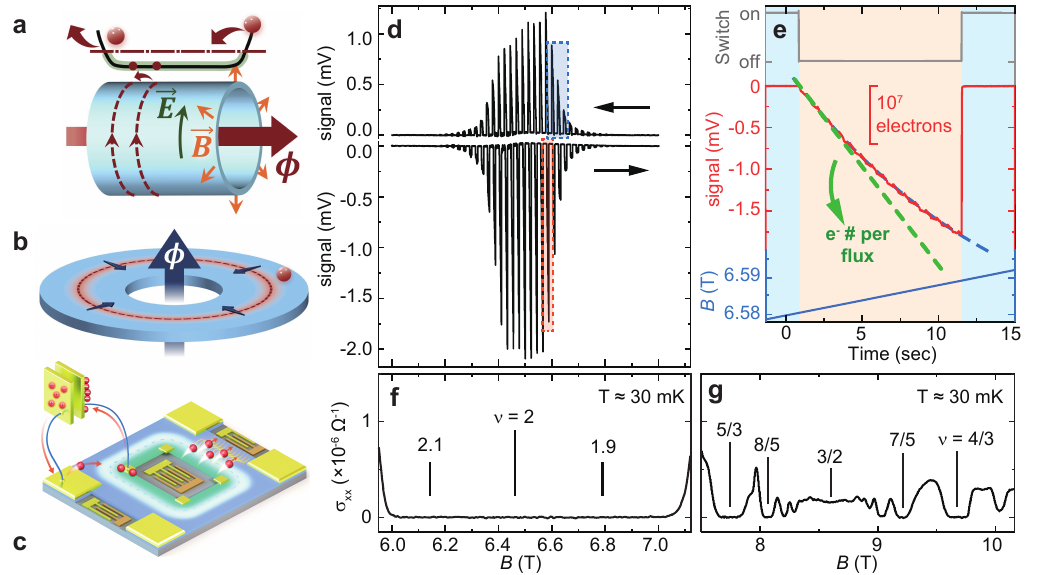}
\caption
{
\textbf{Illustration of the measurement protocol and raw experimental data.} \quad
        (a) Schematic of Laughlin pumping on a cylinder. The magnetic field $\vec{B}$ is along the radial direction and the Landau orbitals are exponentially localized along the axial direction (red dashed curves). One additional flux tube is inserted along the axis of the cylinder. Its strength is adiabatically tuned from zero to $h/e$, which induces an electrical field $\vec{E}$ along the perimeter of the cylinder. Each Landau orbital shifts gradually to its neighbor in this process. If the orbitals carry electrons, an electric current flows from one end of the cylinder to the other. The flowing direction is perpendicular to $\vec{E}$, so the system has a nonzero Hall conductance.
        (b) The Corbino disk version of Laughlin pumping. One flux tube is inserted at the center such that an azimuthal electric field is generated and the Landau orbitals shift along the radial direction. This pumps electrons between the two sections of the Corbino disk. 
        (c) Schematic of our device. When electrons are pumped across the system, they are drained into one capacitor that connects one contact from each section. Its voltage is measured from which the amount of pumped charge is computed. SAWs are launched by three IDTs to clear accumulated charges. 
        (d) Laughlin pumping of the $\nu=2$ integer QHS. The horizontal axis is aligned with the $\nu=2$ plateau in panel (f). We can only observe useful signals in a small region around the plateau center. If the sweeping direction of $B$ is reversed (black arrows), the pumping direction changes accordingly. The data in the orange and blue dashed boxes is presented in Figs.~1c and 3a, respectively.
        (e) One typical measuring cycle. The voltage (red solid curve) increases with the magnetic field $B$. An exponential fitting (blue dashed curve) of the data is performed to extract the pumping rate at zero bias (green dashed line) as well as the longitudinal conductivity. SAWs are switched on periodically to reset the voltage.
        (f \& g) The longitudinal conductivity obtained by conventional transport measurements. We indicate the positions of several QHSs.  
}
\label{Figure1}
\end{figure*}

In two-dimensional electron systems (2DESs), electrons are free to move in a plane but are frozen by quantum confinement along the third direction. The application of a perpendicular magnetic field generates Landau levels (LLs) where a large number of single-particle orbitals have the same kinetic energy. This sets the stage for the quantum Hall states (QHSs)~\cite{Klitzing1980,TsuiDC1982}, where $\sigma_\mathrm{xy}$ attains quantized values in units of $e^{2}/h$ (with $h$ and $e$ being the Planck's constant and elementary electric charge) at the macroscopic scale despite the omnipresence of disorder. To understand this phenomenon, Laughlin analyzed a gedanken process on a cylinder~\cite{Laughlin1981} (see Fig.~\ref{Figure1}a). Along its axis, there is one flux tube whose strength increases adiabatically from zero to $h/e$. An electric field is generated such that a current flows from one end of the cylinder to the other due to nonzero Hall conductance. The total amount of transferred charges is $\sigma_{\mathrm{xy}}h/e$. On the other hand, each Landau orbital moves gradually during flux insertion and eventually reaches the previous location of its neighbor. For integer QHSs with $n$ fully occupied LLs, $n$ electrons are transported between the two ends. This yields the equality $\sigma_{\mathrm{xy}}h/e=ne$ so the Hall conductance is quantized. An intriguing consequence follows when this analysis is applied reversely on fractional QHSs with $\sigma_\mathrm{xy}$ being a fraction $\nu$ times $e^{2}/h$. For each inserted flux, ${\nu}e$ charge is pumped across the cylinder, so that quasi-particles carry fractional charges. 

Laughlin pumping has been an indispensable theoretical tool, but its experimental demonstration is extremely challenging. This process can be reformulated on the Corbino disk, which is an annulus with contacts on inner and outer sections (see Fig.~\ref{Figure1}b)~\cite{Adams1915,DaviesBook}. As one flux tube is inserted at the center, an azimuthal electric field is induced to pump electrons between the two sections. An experimental implementation of such a device is still very difficult, if not impossible. Alternatively, Widom and Clark proposed to study charges accumulated on a floating front gate~\cite{Widom1982}, which was investigated by experiments~\cite{Shashkin1986,Dolgopolov1992,Dolgopolov2011}. However, apparent charge pumping in such systems may coincide with two trivial processes: the electron chemical potential has a discontinuity at integer fillings due to Landau quantization~\cite{DaviesBook,Zeller1986}; the eddy current flowing along the sample perimeter changes its bulk chemical potential according to the Hall voltage~\cite{Huels2004,Klaffs2004}. In short, it is reasonable to say that Laughlin's original vision has not been verified in a satisfactory manner~\cite{Dolgopolov1992,Dolgopolov1993}.

This paper reports a breakthrough that brings Laughlin's gedanken experiment to reality. We fabricate Corbino-type devices with no gates so the energy penalty associated with local charge fluctuations helps to ensure constant areal density $n$. The magnetic field is swept at a constant rate to induce a static azimuthal electric field. DC current is used to avoid AC complications such as the Faraday effect. To determine the amount of pumped charges, they are drained into a capacitor and its voltage $V$ is measured. Since a continuous buildup of $V$ is detrimental, we devise a novel scheme called the sonication of electron accumulation (SEA) to discharge the capacitor in-situ. We demonstrate unambiguously quantized charge pumping in several integral and fractional QHSs. In the presence of finite $V$, the small but nonzero $\sigma_\mathrm{xx}$ reduces the pumping rate. This effect could be harnessed to extract ultralow $\sigma_{\mathrm{xx}}$ beyond the resolution of conventional transport measurements. We obtain effective energy gaps from these data that are drastically different from conventional transport results, suggesting that a completely different mechanism is at work.

\vspace{1em}
\noindent\textbf{Measurement protocol}
\vspace{1em}

The device configuration is displayed in Fig.~\ref{Figure1}c, where the 2DES mesa is topologically equivalent to the Corbino ring. One inner and one outer contact are connected by a capacitor whereas the other contacts are kept floating. Due to the conducting edge channels, all contacts connected to the same edge are in equilibrium~\cite{Ezawa2013,Jain2007,Eisenstein1990,Klitzing2011}. We have studied many combinations of contacts in SI Sec.~VII and the result does not depend on which contact from each section is chosen. The magnetic field is swept slowly at a constant rate $\mathrm{d}B/\mathrm{d}t$ ranging from 0.1 to 3 mT/s. The capacitor drains the pumped electrons from the contacts. Its voltage $V$ builds up linearly with time if the sample realizes a robust QHS. We deduce the amount of pumped charges via $Q=CV$ as shown in Fig.~\ref{Figure1}e, where the capacitance $C$ is measured separately using a capacitor bridge. The direction of the pumping current is dictated by how the magnetic field is swept. For the specific direction in Fig.~\ref{Figure1}e, electrons will be pumped inwards (from outer to inner contacts) with increasing field strength, which leads to a reduction of the voltage $V$. If the field strength decreases, electrons would move in the opposite direction and $V$ increases. We may also reverse the $B$ direction such that positive $\mathrm{d}B/\mathrm{d}t$ corresponds to charge pumping from the inner to outer contacts.

While Laughlin pumping does not involve an electric field along the pumping direction, a finite voltage bias against this direction inevitably appears due to the accumulated charges. Better resolution of the electron number can be achieved using smaller capacitance, but this would lead to a faster buildup of the voltage. As shown in Fig.~\ref{Figure1}e, the measured trace (red solid line) deviates downward from the straight green dashed line, indicating a smaller pumping rate at finite $V$. Although a faster sweeping rate increases the pumping current, continuous increase of $V$ can trigger breakdown of QHSs (see SI Sec.~III). It is not possible to perform reliable measurements without resetting $V$ using an electrically controlled switch with fC-level charge injection and fA-level leaking current. The SEA protocol is developed to address this challenge. We fabricate several inter-digital transducers (IDTs) to launch surface acoustic waves (SAWs)~\cite{White1965,Wixforth1986,Willett1990,Paalanen1992,Esslinger1992,Shilton1995,Simon1996,Shilton1995,Friess2017}. This allows us to tune the longitudinal conductivity $\sigma_\mathrm{xx}$. As the SAW power is raised, $\sigma_\mathrm{xx}$ increases exponentially by several orders of magnitude (see SI Sec.~VIII). In other words, our device can also be interpreted as a SAW-controlled switch with negligible charge injection. During our measurements, SAWs are launched every few seconds to release the charges accumulated on the capacitor. The pumping rate $\mathrm{d}Q/\mathrm{d}t$ can then be extrapolated to $V=0$ by fitting the $V$ vs. $t$ curve.

\vspace{1em}
\noindent\textbf{Precise quantization of the pumping coefficient}
\vspace{1em}

The central quantity of our interest is the pumping coefficient $n_{\Phi}$, defined as the ratio between the number of pumped electrons $Q/e$ and inserted flux $\Phi$ in units of $h/e$. If the QHS is strong, $n_{\Phi}$ should be equal to the Hall conductance $\sigma_{\mathrm{xy}}$ in units of $e^{2}/h$. When the magnetic field is swept, the total amount of fluxes in the system changes at the rate $(\mathrm{d}B/\mathrm{d}t)S$ with $S$ being an effective area. Using the raw $V$ data collected in measurements, the coefficient can be computed as 
\begin{eqnarray}
n_{\Phi} = \frac{hC}{e^{2}S} \left( \frac{\mathrm{d}B}{\mathrm{d}t} \right)^{-1} \left. \frac{\mathrm{d}V}{\mathrm{d}t} \right|_{V=0}
\end{eqnarray}

In contrast to theoretical discussions that employ infinitely thin flux tubes, the whole Corbino ring is penetrated by fluxes. This necessitates numerical simulations to determine $S$. We model the sample as a 2D conductor with $\sigma_{\mathrm{xx}}=0$ and $\sigma_{\mathrm{xy}}={\rm const}$. As the magnetic field changes, an electric field is induced via $\nabla \times \mathbf{E} = \frac{\mathrm{d}\mathbf{B}}{\mathrm{d}t}$, which leads to the current density $\mathbf{J}=\mathop{\sigma} \limits ^{\leftrightarrow}\mathbf{E}$. We require that the charge density remains constant and the two edges are equipotential lines (see SI Sec.~IV). The current density is integrated to yield the total current $I$ such that
\begin{eqnarray}
S = \frac{h}{e^2\sigma_{\mathrm{xy}}} \left( \frac{\mathrm{d}B}{\mathrm{d}t} \right)^{-1} I.
\end{eqnarray}
For an ideal Corbino ring with inner (outer) radius $r_{1}$ ($r_{2}$), our results are consistent with $S = \frac{\pi}{2}\frac{r^{2}_{2}-r^{2}_{1}}{\ln(r_{2}/r_{1})}$ found previously~\cite{Fontein1988,Kawamura2023}. 

\begin{figure*}[!htbp]
\includegraphics[width=\textwidth]{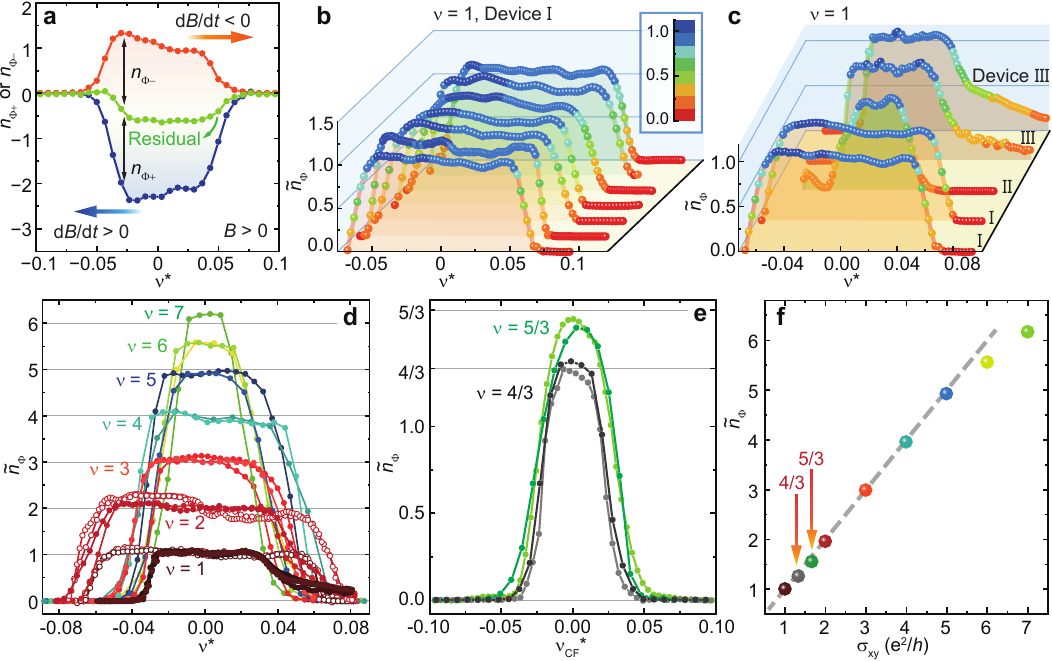}
\caption
{\textbf{Accurate determination of the pumping coefficient.} \quad
        (a) For up and down sweeps in Fig.~\ref{Figure1}e, we obtain the pumping coefficients $n_{\Phi-}$ and $n_{\Phi+}$ (orange and blue curves), respectively. Their difference is $2\tilde{n}_{\Phi}$ and their average (green curve) reflects non-ideal effects in our measurements.
        (b) Quantized $\tilde{n}_{\Phi}$ on the $\nu=1$ plateau in Device I. The seven data sets correspond to different $|dB/dt|$ ranging from 0.05 to 0.1 T/min.
        (c) Comparison of $\tilde{n}_{\Phi}$ in Device I, II, and III. To discharge the capacitor properly, we use different SAW power (-35 or -40 dBm) and switching frequency (from 17.37 to 77 mHz) in panels (b) and (c).
        (d) $\tilde{n}_{\Phi}$ on several integer plateaus in Device I (solid symbol) and III (open symbol). The horizontal axis in panels (a-d) is $\nu^{*}$ defined in the main text.
        (e) $\tilde{n}_{\Phi}$ on the $\nu=4/3$ and $5/3$ plateaus in Device III with horizontal axis being $\nu^{*}_{\rm CF}$ (see SI Sec.~VI).
        (f) Comparison of $\tilde{n}_{\Phi}$ and $\sigma_\mathrm{xy}$. The data points are plateau heights averaged over all traces, which match the expected $\tilde{n}_{\Phi}=\sigma_\mathrm{xy}$ (gray dashed line). The error bar is smaller than the size of the symbol. 
}
\label{Figure2}
\end{figure*}

As illustrated in Fig.~\ref{Figure1}d, we have performed measurements for positive and negative $\frac{\mathrm{d}B}{\mathrm{d}t}$ (referred to as up- and down-sweep). The pumping coefficients are denoted as $n_{\Phi+}$ and $n_{\Phi-}$ as shown in Fig.~\ref{Figure2}a. The results apparently deviate from the expected theoretical values. This is because real-world experiments have imperfections despite our best efforts in the design. We mention some important issues here and relegate the details to SI Sec.~V. If the local chemical potentials vary with the magnetic field, electrons may be driven into or out of the 2DES. However, no net difference between the current passing through the two sections is observed, so the density of electrons should be conserved. The measured $V$ may have a constant offset due to the Faraday effect because the measurement wires enclose a finite area comparable to the sample size (it is the main artifact in AC measurement~\cite{Syphers1986,Fontein1988,Jeanneret1994,Jeanneret1997,Kawamura2023}). Its contribution should be negligible compared to the pumping current in view of the extremely low $\sigma_{\mathrm{xx}}$. The dominant artifact in our experiment is the residual zero offset current of the electrometer. Fortunately, its impact does not depend on the sign of $\mathrm{d}B/\mathrm{d}t$ so it can be eliminated if we define the antisymmetric combination
\begin{eqnarray}
\widetilde{n}_{\Phi} = \frac{1}{2} \left( n_{\Phi-}-n_{\Phi+} \right)
\end{eqnarray}
as the pumping coefficient.

The measured values of $\widetilde{n}_{\Phi}$ for the $\nu=1$ QHS from Device I are presented in Fig.~\ref{Figure2}b. The seven sets of data correspond to different choices of $\frac{\mathrm{d}B}{\mathrm{d}t}$. In all cases, plateaus with approximately unity height were found in the range of $\nu^{*}=\nu-1 \in [-0.05,0.05]$. It is considerably narrower than the plateau observed by conventional measurements (see SI Sec.~II). As an intrinsic property of QHS, $\widetilde{n}_{\Phi}$ should not depend on the system size as long as it is sufficiently large. This anticipation is confirmed in Fig.~\ref{Figure2}c, where data from other devices exhibit plateaus with narrower widths but the same height. We have studied many integer QHSs and the results are summarized in Fig.~\ref{Figure2}d. Clearly resolved $\widetilde{n}_{\Phi}$ plateaus were observed for $\nu$ up to $7$. The QHSs become weaker as the filling factor increases, so the plateau widths shrink and their heights gradually deviate from the expected values. If we reduce $\mathrm{d}B/\mathrm{d}t$ to 0.3 mT/s, the phase coherence length 
\begin{eqnarray}
L=\ell_{B}\left(\frac{2\pi \omega_{c} B}{\mathrm{d}B/\mathrm{d}t} \right)^{1/4}\approx 0.4 \, \mathrm{mm}
\end{eqnarray}
would be comparable to the sample size ($\omega_{c}$ is the cyclotron frequency and $\ell_{B}$ is the magnetic length), thus the adiabatic limit is reached. To visualize the data, we summarize the averaged values of $\widetilde{n}_{\mathrm{\Phi}}$ versus $\sigma_\mathrm{xy}$ in Fig.~\ref{Figure2}f. The points fall nicely on a straight line except for the largest two $\nu$. By fitting the strong QHSs at $\nu \leq 5$, we find a remarkable accuracy of $\tilde{n}_\Phi/\sigma_\mathrm{xy}=0.99\pm 0.02$, including all measurement uncertainties. 

When Laughlin pumping is applied to fractional QHSs, an even more intricate property is revealed. Instead of trying to derive the Hall conductance, we accept fractionally quantized Hall conductance as an experimental fact and explore its consequences. It is immediately clear that adiabatic insertion of a flux tube would pump a fractional charge. This is a powerful conclusion that does not rely on the microscopic origins of fractional QHSs. Historically, Laughlin found an elegant wave function for the $1/3$ state which unveiled charge-$e/3$ elementary excitations~\cite{Laughlin1983}. Experimental investigations detected fractional charges using shot noise~\cite{Picciotto1997,Saminadayar1997}, scanning single electron transistor~\cite{Martin2004}, and Coulomb blockade~\cite{Roosli2021}. However, pumping of fractional charges has not been demonstrated to date. An important challenge is the small magnitude of energy gaps in fractional QHSs and the finite residual longitudinal conductivity. Using our protocol, we make an essential step forward. Due to the limited range of magnetic field, we study the $\nu=4/3$ and $5/3$ states. The integer parts form integer QHSs and the fractional parts form fractional QHSs at $1/3$ and $2/3$ fillings. When one flux tube is inserted, $4e/3$ and $5e/3$ charges are pumped in these states, respectively. As displayed in Fig.~\ref{Figure2}e, $\widetilde{n}_{\Phi}$ for them are quite close to the anticipated values. This match appears in a very small window since the fractional QHSs are much weaker than the integer ones. We have made use of the composite fermion (CF) theory to label the horizontal axis~\cite{Jain1989}. In this framework, one electron is dressed with two fluxes to become CFs that experience an effective magnetic field. The CFs can be organized into effective LLs with filling factor $\nu_{\rm CF}$. For the $1/3$ ($2/3$) part, we have $\nu_{\rm CF}=1$ ($\nu_{\rm CF}=2$) and the effective magnetic field is parallel (opposite) to the actual magnetic field. The deviation of the CF filling factor from integer values is defined as $\nu^{*}_{\rm CF}$ (see SI Sec.~VI). 

\vspace{1em}
\noindent\textbf{Ultralow longitudinal conductance}
\vspace{1em}

In realistic settings, QHSs have finite bulk longitudinal conductivity $\sigma_{\mathrm{xx}}$, whose magnitude may be used to characterize robustness of the states. For the Hall bar geometry, the longitudinal and Hall resistances are determined by measuring the voltage drops at fixed input current. It is not suitable to deduce $\sigma_{\mathrm{xx}}$ from these results because the current generally has a non-uniform distribution inside the sample and exhibits complicated evolution across one plateau~\cite{Klitzing2011,WuMengmeng2024}. The Corbino geometry is preferable for probing $\sigma_{\mathrm{xx}}$ in conventional transport measurements. In such cases, a voltage bias is applied between the two sections and the flowing current is measured. Its resolution is usually limited to $\sim 10^{-9}\, \Omega^{-1}$. To observe Laughlin pumping, $\sigma_{\mathrm{xx}}$ should be suppressed so the back leaking current is small. Meanwhile, the leaking current also provides a new way to determine $\sigma_{\mathrm{xx}}$ that is well beyond the reach of transport measurements. The linear charging curve in Fig.~\ref{Figure1}c saturates as charge accumulates on the contacts due to finite $\sigma_{\mathrm{xx}}$. We can fit the data using an exponential function $V_{0}[1-\exp(-t/\tau)]$ such that $\sigma_{\mathrm{xx}}=\beta C/\tau$ with $\beta$ being a geometry dependent factor given by numerical simulation (see SI Sec.~IV). The minimal value of $\sigma_{\mathrm{xx}}$ that we have measured is about $10^{-13}\, \Omega^{-1}$, which leads to settling time $\tau$ up to ten hours. If AC measurements were performed using a metallic front gate, the current would be mostly confined to a ``skin layer" whose width is about $10$ \textmu m at 2 Hz~\cite{Dolgopolov2011}.

\begin{figure}[!htbp]
\includegraphics[width=0.67\textwidth]{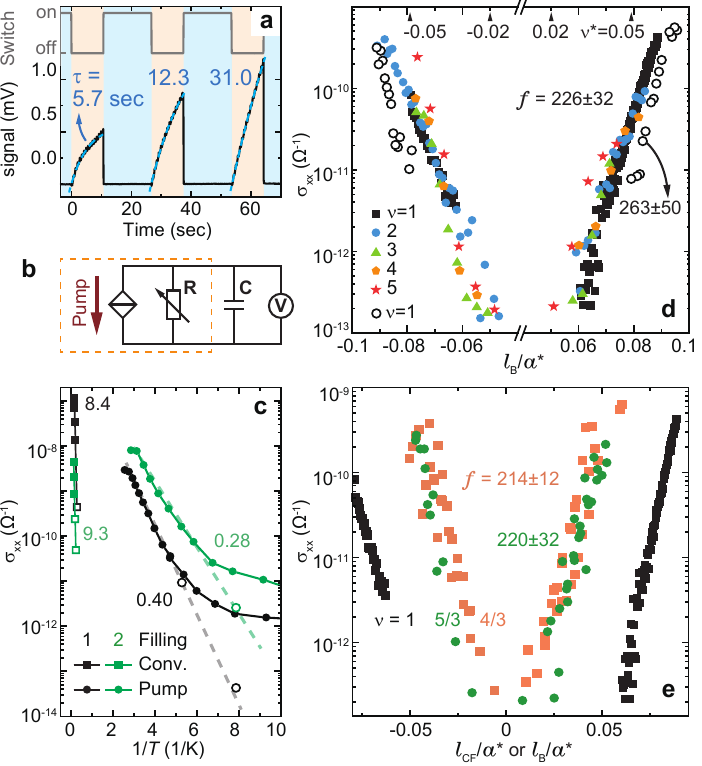}
\caption
{\textbf{Extraction of minuscule $\sigma_\mathrm{xx}$ on quantum Hall plateaus.} \quad
        (a) Zoom-in view of the blue dashed box in Fig.~\ref{Figure1}e. 
        (b) Each curved segment in panel (a) is modeled by a circuit that consists of a leaking resistance $R$ in parallel with the capacitor $C$. We can extract the time constant $\tau=RC$ by fitting each segment and then convert $R$ to the longitudinal conductivity $\sigma_\mathrm{xx}$.
        (c) $\sigma_\mathrm{xx}$ vs. $1/T$ at exactly $\nu=1$ and $2$. The solid circles (squares) are obtained by our new method (conventional transport measurements). The open symbols are extrapolated values at $\nu^*=0$ (see SI Sec.~VII). Arrhenius fitting is performed to extract the activation gaps that are displayed in the vicinity of each data set (in units of meV).
        (d-e) The evolution of $\sigma_\mathrm{xx}$ on several quantum Hall plateaus when the filling factor varies. For integer (fractional) QHSs, the horizontal axis is $\ell_{B}/\alpha^{*}$ ($\ell_{\rm CF}/\alpha^{*}$) (see SI Sec.~VI). The solid (open) symbols in panel (d) are from Device I (IV). The fitting parameter $f$ defined in the main text are indicated for different data sets.
}
\label{Figure3}
\end{figure}

We plot in Fig.~\ref{Figure3}c the extracted $\sigma_{\mathrm{xx}}$ at exact integer fillings $\nu=1$ and 2. For comparison, we also present conventional transport results which are only feasible at temperatures above $\sim 2$ K. We may define a pumping activation gap $\Delta$ by fitting $\sigma_{\mathrm{xx}}$ as $\exp[-\Delta/(2k_{B}T)]$. Its value is $0.40$ meV ($4.6$ K) at $\nu=1$ and $0.28$ meV ($3.2$ K) at $\nu=2$, which are much smaller than the conventional activation gap deduced from transport data [$8.4$ meV ($98$ K) for $\nu=1$ and $9.3$ meV ($108$ K) for $\nu=2$]. This finding points to the existence of more than one mechanism underlying charge transport. To further unveil the properties of plateaus, we explore the vicinity of exact integer fillings. As shown in Fig.~\ref{Figure3}d, the magnitude of $\sigma_\mathrm{xx}$ on several integer plateaus grows by about three orders of magnitude when $\nu^{*}$ changes from $0.02$ to $0.05$ (data taken at $\sim 30$ mK). This rise is much steeper than any polynomial dependence predicted by the variable range hopping~\cite{Mott2012}. 


Although a quantitative explanation is not known, data analysis can provide some hints. Some excitations are created at finite $\nu^{*}$: electrons are pushed into higher LLs that were empty or orbitals are vacated in LLs that were fully occupied. The average distance between these excitations is defined as $a^{*}$ (see SI Sec.~VI). If we choose the horizontal axis to be $\ell_{B}/a^{*}$, the data points for different QHSs in Device I (solid symbols) coalesce on roughly the same straight line, i.e. $\sigma_\mathrm{xx} \propto \exp (f\ell_{B}/a^{*})$ with one fitting parameter $f\approx 226$ (more data is given in SI Sec.~VIII). It becomes more interesting as the data taken from Device IV (open symbols in Fig.~\ref{Figure3}d) exhibit the same dependence with $f\approx 263$. This is quite surprising because its density is halved and its mobility is nearly ten times smaller than that of Device I, and their conventional transport results are conspicuously different (see SI Sec.~II). We tentatively link this behavior with the formation of disorder-pinned Wigner crystals~\cite{ChenYong2003,Myers2024}. Thermal fluctuations generate defects to carry a small current. As the average distance $a^{*}$ increases, the crystals become more compact and defects are more likely to proliferate, which gives rise to larger $\sigma_{\mathrm{xx}}$. An important consequence is that the conductivity does not depend on many specific properties of the QHSs. This scenario is consistent with the fact that $\sigma_{\mathrm{xx}}$ is symmetric in $\nu^*$ and exhibits the same trend in different QHSs. For the fractional plateaus at $4/3$ and $5/3$, we replace the horizontal axis by $\ell_{\rm CF}/a^*$ with $\ell_{\rm CF}$ being the magnetic length of CFs (see SI Sec.~VI). It is remarkable that $\sigma_{\mathrm{xx}}$ can still be fitted as $\exp (f\ell_{\rm CF}/a^{*})$ using approximately the {\em same} $f$. We speculate that this is due to the presence of CF Wigner crystals~\cite{ZhuHan2010,Archer2013}. 

\vspace{1em}
\noindent\textbf{Conclusions}
\vspace{1em}

In conclusion, we have demonstrated Laughlin pumping in multiple QHSs. Our experimental protocol is very different from previous ones. An essential ingredient is the SEA method, which periodically clears the charges accumulated on the capacitor. Thanks to this reset of the capacitor voltage, the pumping coefficient $\tilde{n}_{\rm \Phi}$ at zero bias can be extracted and well-quantized results are obtained for several QHSs. Most notably, we measure directly fractional charge pumping for two fractional QHSs. Furthermore, we unveil ultralow longitudinal conductivity of QHSs beyond the resolution of conventional transport measurements. This provides a glimpse of the intricate internal structures of plateaus that call for more in-depth studies.

Many interesting directions may be explored in future work. As demonstrated above, quantized charge pumping can only be achieved when the longitudinal conductivity is sufficiently small. This makes it a challenge to study more fragile states such as the putative non-Abelian one at $5/2$~\cite{Willett1987,MaKW2024}. We hope to address this challenge with better experimental design and higher sample quality. To realize Laughlin pumping, the parameters should be varied slowly to make sure that the system evolves adiabatically. If this constraint is relaxed, the response is unclear. Generally speaking, nonequilibrium phenomena in QHSs are largely unexplored. In recent years, remarkable progress has been made on quantum Hall physics without magnetic fields, namely fractional Chern insulators or fractional quantum anomalous Hall effect~\cite{Bernevig2025}. It is natural to ask how fractional charge pumping can be demonstrated in these cases.


\bibliography{LaughlinPump}

\vspace{1em}
\noindent\textbf{Acknowledgments}
\vspace{1em}

We acknowledge support by the National Key Research and Development Program of China (Grant No. 2021YFA1401900), the Quantum Science and Technology-National Science and Technology Major Project (Grant No. 2021ZD0302602), and the National Natural Science Foundation of China (Grants No. 12574184, No. 12450003, and No. 12174130). The Princeton University portion of this research is funded in part by the Gordon and Betty Moore Foundation’s EPiQS Initiative, Grant No. GBMF9615.01 to Loren Pfeiffer. 


\vspace{1em}
\noindent\textbf{Author contributions}
\vspace{1em}

Y.L. conceived and supervised the study. R.W. and X.Liu designed the fabrication processes. R.W. performed the pumping measurements. X.Liu fabricated the devices. A.G. grew wafers A and B with assistance from K.W.B. and under the supervision of L.N.P. W.Z. grew wafer C under the supervision of R.R.D. R.W., Y.W. and Y.L. analyzed the data and wrote the manuscript with inputs from M.S. and X.Lin. 



\vspace{1em}
\noindent\textbf{Competing interests}
\vspace{1em}

The authors declare no competing interests.

\clearpage

\clearpage

\setcounter{figure}{0}
\setcounter{table}{0}
\setcounter{equation}{0}
\renewcommand*{\thefigure}{S\arabic{figure}}
\renewcommand{\thetable}{S\arabic{table}}
\renewcommand{\theequation}{S\arabic{equation}}

\begin{center}
	\noindent\textbf{Supplementary Information for "Laughlin pumping assisted by surface acoustic waves"}
\end{center}
\tableofcontents

\section{Methods and Samples}

\begin{figure}[!htbp]
	\includegraphics[width=1\textwidth]{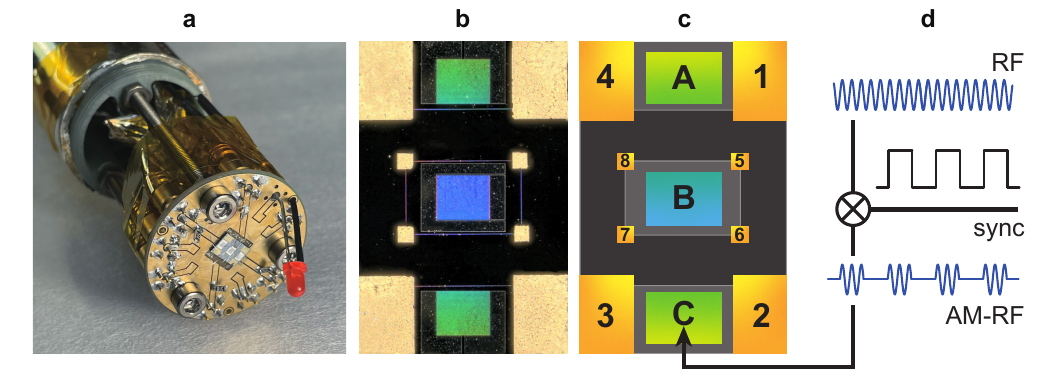}
	\caption{
         (a) Device I installed on the sample holder and the cold-finger.
         (b,c) Sample photo and schematic diagrams. The 2DES mesa has square Corbino geometry with eight contacts (labeled 1–8) and three IDTs (labeled A–C). 
         (d) An on-off modulated (at about 100 mHz) excitation RF signal is applied to the emitting IDT used to excite the SAW. 
		}
  \label{figS1} 
\end{figure}

Our GaAs/AlGaAs samples are made from three different wafers (A, B and C) grown by molecular beam epitaxy. The wafers A and B were grown at Princeton University, and the wafer C was grown at Peking University. We have studied 4 devices (I, II, III and IV) and their detailed information is summarized in Tab.~\ref{Tab_5_SampleList}. Figure~\ref{figS1}(a) and (b) show the photo of our sample holder and Device I, respectively. All samples were fabricated using standard photolithography and wet etching. The two-dimensional electron system (2DES) mesa has a rectangular Corbino geometry with eight evaporated Au/Ge/Ni/Au contacts. The gold squares labeled by 1-4 are the outer contacts and those labeled by 5-8 are inner contacts. Three 5 \textmu m-period interdigital transducers (IDTs) are evaporated outside the 2DES mesa. The IDT A and C are located outside the square ring of the 2DES, and the IDT B locates inside the ring. All doping layers underneath the IDTs are removed by wet etching. All contacts are placed outside the propagation path of surface acoustic wave (SAW) to reduce unnecessary disturbances. 

During the experiment, all contacts are connected to coaxial cables such that the leaking conductance is suppressed to a level below $10^{-12} \, \Omega^{-1}$. We use the sub-nF cable capacitor to drain the pumped charge, and monitor its voltage using a compact battery powered charge sensor with $>1$ T$\Omega$ input impedance. The capacitor value is measured precisely by a capacitance bridge. We have achieved charge resolution that is well below fC, and the piezo charge induced by the pulse tube vibration is our major limitation (see SI Sec.~V). Conventional transport measurements are performed using standard lock-in technique (see SI Sec.~II). We use a radio frequency (RF) source at frequency $590$ MHz and power $P_\mathrm{in} = -30 \sim -50$ dBm to excite the IDTs. We placed a 50$\Omega$ resistor in parallel with the IDT for broadband impedance matching. The typical insertion loss between a pair of these IDTs is about 30-35 dB, indicating that the SAW power $P_{\rm SAW}$ is less than 1\% of the input power $P_{\rm in}$. The switching of acoustic waves is achieved by amplitude modulating this RF signal output at proper modulation frequency, e.g. $10 \sim 500$ mHz. All experiments are carried out in a dilution refrigerator whose base temperature is below 10 mK.

\begin{table}[h]
	\centering
	\renewcommand{\arraystretch}{1.5} 
	\label{Tab_5_SampleList}
	\begin{tabular}{c|cccccccc}
	\hline	
  Sample  & Wafer   & Structure      & \begin{tabular}[c]{@{}c@{}}Density\\ ($\times 10^{11} $ cm$^{-2}$)\end{tabular} & \begin{tabular}[c]{@{}c@{}}Mobility\\ (cm$^2$/(V$\cdot$s))\end{tabular} & \multicolumn{1}{c}{\begin{tabular}[c]{@{}c@{}}Outer square \\ (mm)\end{tabular}} & \multicolumn{1}{c}{\begin{tabular}[c]{@{}c@{}}Inner square \\ (mm)\end{tabular}} & \multicolumn{1}{c}{\begin{tabular}[c]{@{}c@{}}Effective area S \\ (cm$^2$)\end{tabular}} & $\beta=\sigma_\mathrm{xx}/G$ \\ \hline \hline
		Device I   & A & 30-nm QW  & $3.0 \times 10^{11} $                                                      & $>2 \times 10^7$                                                      & $4.0 \times 3.4$                                                         & $2.2 \times 1.4$     & 0.060 & 0.103
                                                   \\
		Device II  & B & 30-nm QW  & $3.0 \times 10^{11} $                                                      & $>2 \times 10^7$                                                      & $4.0 \times 2.2$                                                         & $3.0 \times 1.5$    & 0.045 & 0.039
                                         \\
		Device III & B & 30-nm QW  & $3.0 \times 10^{11} $                                                     &  $>2 \times 10^7$                                                    & $4.0 \times 1.0$                                                         & $2.8 \times 0.4$      & 0.019    & 0.044                                           \\
		Device IV  & C & heterostructure &  $1.2 \times 10^{11} $                                                   & $>1 \times 10^6$                                                      & $4.0 \times 3.4$                                                         & $2.2 \times 1.4$   & 0.060 & 0.103  \\ \hline
	\end{tabular}
 \caption{The list of devices used in this study.}
\end{table}

\section{Conventional transport results}

\begin{figure}[!htbp]
	\includegraphics[width=\textwidth]{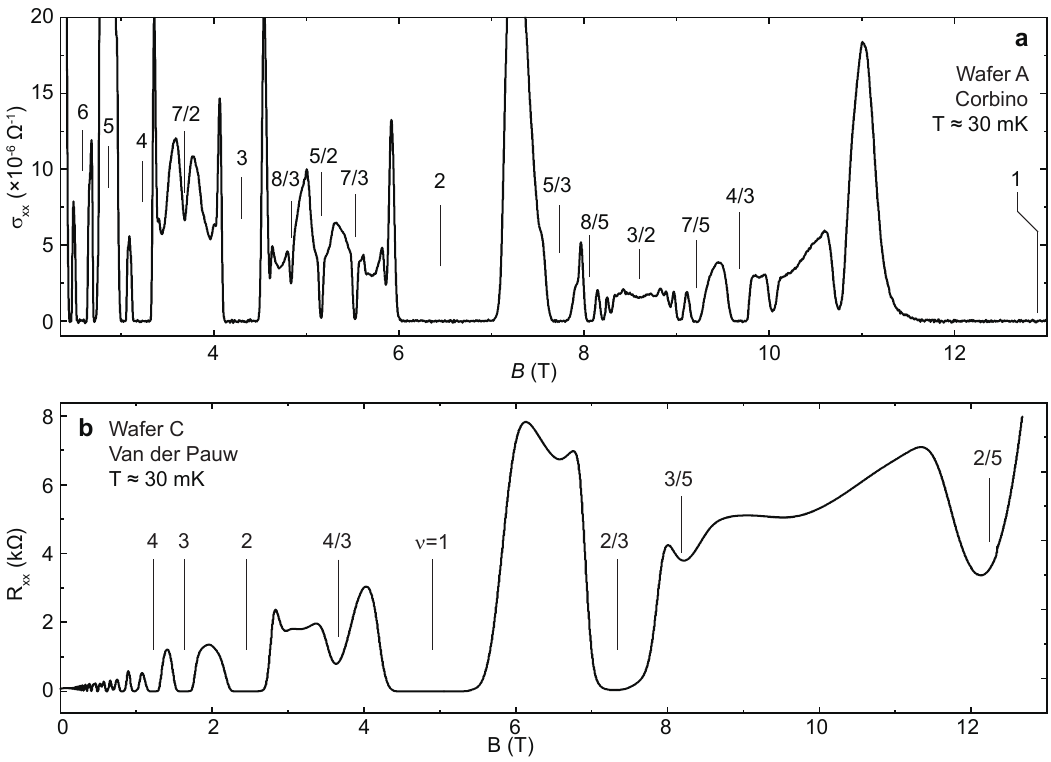}
	\caption{Conventional quasi-DC transport measurement in the absence of the SAW.
            (a) Longitudinal magneto-conductance $\sigma_\mathrm{xx}$ taken from a Corbino sample made from wafer A. The sample has a standard circular Corbino geometry with the inner and outer radii $r_1=400$ \textmu m, $r_2=600$ \textmu m, respectively. We apply a voltage of 10 \textmu V between inner and outer contacts \textmu V and measure the current flowing through the sample. (b) Longitudinal magneto-resistance $R_\mathrm{xx}$ taken from a Van der Pauw geometry sample (device IV) made from wafer C. 
		}
  \label{Fig_S_Transport} 
\end{figure}

Figure~\ref{Fig_S_Transport} shows the magneto-transport results measured from wafers A and C using the conventional lock-in technique. Fig.~\ref{Fig_S_Transport}(a) is the longitudinal conductivity $\sigma_\mathrm{xx}$ measured using Corbino geometry. The sample shown in Fig.~\ref{Fig_S_Transport}(a) is made from the wafer piece adjacent to the Device \uppercase\expandafter{\romannumeral 1}. The measured conductance $G$ is converted to the conductivity $\sigma_\mathrm{xx}$ using a geometry factor $\beta$ defined by Eq.~\eqref{eq_transport}. 
\begin{equation}\label{eq_transport}
	\sigma_\mathrm{xx} = \frac{\ln(r_2/r_1)}{2\pi} G = \beta G
\end{equation}
The integer quantum Hall (IQH) effects, as well as strong fractional quantum Hall (FQH) effects at $\nu=4/3$ and 5/3, exhibit well-developed zero-current plateaus. A series of FQH effects emerge, including the fragile $\nu=$5/2 state. Meanwhile, Fig.~\ref{Fig_S_Transport}(b) shows the longitudinal resistivity from a Hall bar geometry sample made from wafer C. The mobility of this wafer is ten times smaller than wafers A and B. The IQH effects are strong, but $R_{\rm xx}$ does not reach zero even at the strongest FQH effects at $\nu=2/3$, 4/3, etc.

\section{Charge pumping without SEA}

Early attempts of charge pumping experiment \cite{Pudalov1984,Shashkin1986,Dolgopolov1990, Dolgopolov1991,Dolgopolov1992,Lee1997,Dolgopolov2001,Dolgopolov2011} observe the voltage drop across the Corbino ring sample built up by the dc-varying magnetic field. The slope of the voltage with respect to the magnetic field is related to the Hall conductivity through Eq.~\eqref{eq_5_DC_QT}.
\begin{equation}\label{eq_5_DC_QT}
	C \frac{d V}{dB} = \sigma_\mathrm{xy} S(r_1,r_2),
\end{equation}
where the effective area $S(r_1,r_2)$ is related to the sample's geometry. The pumping direction reverses if the magnetic field sweep direction flips (e.g. up sweep to down sweep), so that the voltage polarity changes accordingly. 

\begin{figure}[!htbp]
	\includegraphics[width=1\textwidth]{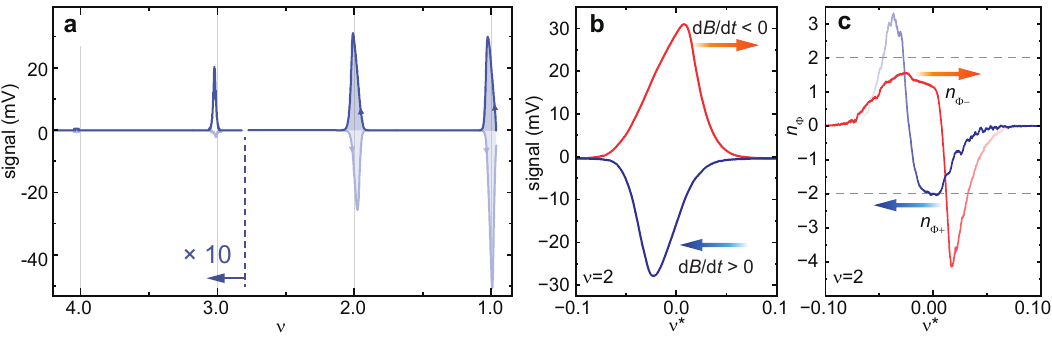}
	\caption{\label{Fig_S_DC_AC_Methods} 
		The charge pumping experiment performed on the Device \uppercase\expandafter{\romannumeral 1} without SEA. (a) The measured capacitance voltage signal at a constant magnetic field sweeping rate of 0.1 T/min. The sweeping direction is marked by arrows on the curves. Note that the signals at $\nu>2.5$ is amplified by a factor of 10 for clearer visualization. (b) Enlarged signal near $\nu=2$ exhibit a parallelogram hysteretic loop. (c) Deduced $n_{\rm \Phi\pm}$ by numerical derivation of the signal. 
	}
\end{figure}

Figure~\ref{Fig_S_DC_AC_Methods} repeats the same measurement on the Device \uppercase\expandafter{\romannumeral 1} sample. The results are consistent with earlier reports. Inside the IQH plateau, the charges are pumped and accumulated at the inner and outer contacts if the magnetic field is swept with a constant rate. However, without our SEA method, the continuous increase in voltage will cause the charge accumulation rate to deviate from its zero-bias value, making it impossible for quantitative investigation. Meanwhile, the excessively high voltage between inner and outer contacts will trigger the breakdown of the quantum Hall effect even when $\nu^*$ is very small. The signal exhibits a parallelogram hysteretic loop in Fig.~\ref{Fig_S_DC_AC_Methods}(b). We can deduce $n_{\rm \Phi\pm}$ by numerical derivation of the Fig.~\ref{Fig_S_DC_AC_Methods}(b) data. However, the resulting Fig.~\ref{Fig_S_DC_AC_Methods}(c) is consistent but less conclusive compared to Fig.~2 of the main text. The plateau-like structure in $n_{\rm \Phi\pm}$ barely develops for a very narrow range of $\nu^*$ before the quantum Hall breakdown. Because of the finite $\sigma_{\rm xx}$, the leaking current can be comparable to the pumping current when the voltage bias is large. In short, it is impossible to complete a quantitative measurement of $n_{\rm \Phi}$ without SEA.

\section{Determination of effective sample area}

\subsection{Circularly symmetric sample}

Consider a circularly symmetric annular sample with inner radius of $r_1$ and outer radius of $r_2$. The conductivity tensor in the polar coordinate is defined by Eq.~\eqref{eq_5_Cal_Sigma},
\begin{equation}\label{eq_5_Cal_Sigma}
	\mathop{\sigma} \limits ^{\leftrightarrow} =
	\begin{pmatrix}
		\sigma_\mathrm{rr} 		 & \sigma_\mathrm{r \theta} \\
		\sigma_\mathrm{\theta r} & \sigma_\mathrm{\theta \theta} \\
	\end{pmatrix}
\end{equation}
where $\sigma_\mathrm{\theta r} = -\sigma_\mathrm{r \theta} $. According to Ohm's Law, 
\begin{equation}\label{eq_5_Cal_Ohms}
	\begin{pmatrix}
		j_\mathrm{r}  \\ 
		j_\mathrm{\theta}
	\end{pmatrix} 
	=
	\begin{pmatrix}
		\sigma_\mathrm{rr} 		 & \sigma_\mathrm{r \theta} \\
		\sigma_\mathrm{\theta r} & \sigma_\mathrm{\theta \theta} \\
	\end{pmatrix}
	\begin{pmatrix}
		E_\mathrm{r}  \\ 
		E_\mathrm{\theta}
	\end{pmatrix} 
\end{equation}
We can assume zero charge accumulation inside the sample
\begin{equation}\label{eq_5_dndt}
	e^{-1}\frac{\mathrm{d}n}{\mathrm{d}t}=\bm{\nabla}\cdot\bm{j}=0 \rm ,
\end{equation}
where $n$ is the electron density. The pump current density $\bm{j}$ is circular symmetric and the total current $I_\mathrm{p}= 2\pi r j_\mathrm{r}$. Meanwhile, according to the configuration of the external measurement circuit, $I_\mathrm{p}$ is related to the voltage $V$ between the inner and outer edge through the boundary condition
\begin{equation}\label{eq_5_Cal_I_CV}
	I_\mathrm{p} = \frac{\mathrm{d}Q}{\mathrm{d}t} = C \frac{\mathrm{d}V}{\mathrm{d}t}
\end{equation}
According to Faraday's Law, the azimuthal electric field can be expressed as Eq.~\eqref{eq_5_Cal_E_theta}.
\begin{equation}\label{eq_5_Cal_E_theta}
	E_\mathrm{\theta} = \frac{r}{2}\left( -\frac{\mathrm{d}B}{\mathrm{d}t}\right),
\end{equation}
where $\mathrm{d}B/\mathrm{d}t$ is a constant in our experiment. The radial electric field can then be deduced as Eq.~\eqref{eq_5_Cal_E_r}.
\begin{equation}\label{eq_5_Cal_E_r}
	\begin{aligned}
		E_\mathrm{r}&= \frac{1}{\sigma_\mathrm{rr}} \left[j_\mathrm{r} - \sigma_\mathrm{r \theta}E_\mathrm{\theta} \right]\\
					&= \frac{1}{\sigma_\mathrm{rr}} \left[\frac{C}{2\pi r}\partial_t V - \sigma_\mathrm{r \theta} \frac{r}{2} \left( -\frac{\mathrm{d}B}{\mathrm{d}t}\right) \right]
	\end{aligned}
\end{equation}
Thus, the measured voltage $V$ can be deduced from the integral of the radial electric field given by Eq.~\eqref{eq_5_Cal_V}.
\begin{equation}\label{eq_5_Cal_V}
	\begin{aligned}
		-V &= \int_{r_2}^{r_1} E_\mathrm{r}\,\mathrm{d}r\\
		& = \frac{C}{2\pi\sigma_\mathrm{rr}} \ln{\frac{r_2}{r_1}}  \left(\partial_t V\right)
			- \frac{\sigma_\mathrm{r \theta}}{4\sigma_\mathrm{rr}} \left(r_2^2 - r_1^2\right) \left( -\frac{\mathrm{d}B}{\mathrm{d}t}\right)
	\end{aligned}
\end{equation}
This equation has a simple form $-V = \tau\partial_t V-\mathrm{A}$ if we define the time constant $\tau$ and parameter $A$ as
\begin{equation}\label{eq_5_Cal_V_solve}
		\tau = \frac{C}{2\pi\sigma_\mathrm{rr}} \ln{\frac{r_2}{r_1}} = \frac{C}{\beta\sigma_\mathrm{rr}},\quad 
		\mathrm{A} = \frac{\sigma_\mathrm{r \theta}}{4\sigma_\mathrm{rr}} \left(r_2^2 - r_1^2\right),
\end{equation}
and its general solution is $V(t)= -\mathrm{A}\exp(-t/\tau) + \mathrm{A}$ if we set the initial condition $V(t=0) = 0$. The full solution can be expressed as Eq.~\eqref{eq_5_Cal_V_solution}.
\begin{equation}\label{eq_5_Cal_V_solution}
	\begin{aligned}
		V(t) &= -\mathrm{A}\exp(-t/\tau) + \mathrm{A} \\
			&= \frac{\sigma_\mathrm{r\theta}}{4 \sigma_\mathrm{rr}}\left(r_2^2 - r_1^2\right) \left( -\frac{\mathrm{d}B}{\mathrm{d}t}\right)
			\left[ 1 - \exp\left( -\frac{t}{(C/2\pi\sigma_\mathrm{rr}) \ln(r_2/r_1)} \right) \right]
	\end{aligned}
\end{equation}
Substituting this solution into Eq.~\eqref{eq_5_Cal_I_CV}, the pump current $I_\mathrm{p}$ at zero bias is 
\begin{equation}\label{eq_5_Cal_Ip_solution}
	I_\mathrm{p}(V=0) = C V^{\prime}(t=0) 
	= \sigma_\mathrm{r\theta} S \left(-\mathrm{d}B/\mathrm{d}t\right)
\end{equation}
and the effective area $S$ of a standard Corbino ring is
\begin{equation}\label{eq_5_AC_f_S_withoutGate}
    S_\mathrm{Ring}(r_1,r_2) = \frac{\pi}{2} \frac{(r_2^2-r_1^2)}{\ln(r_2/r_1)}.
\end{equation}

\subsection{Our square-shaped Corbino devices}

The effective area $S$ of our square-shape Corbino sample, defined according to Eq.~\eqref{eq_5_Cal_Ip_solution}, is obtained using the finite element analysis (FEA) of COMSOL software. The electric field satisfies the Maxwell equations Eqs.~\eqref{eq_5_Cal_E_theta} and \eqref{eq_5_Cal_E_r}. We force charge conservation inside the sample, i.e. Eq.~\eqref{eq_5_dndt}. The boundary condition is set as the following. Each of the inner and outer rings is at equipotential. They are connected via an external capacitor, so that $I_p$ and $V$ satisfy Eq.~\eqref{eq_5_Cal_I_CV}. We divide the system into a triangular mesh, initiate their potential to zero, and simulate the time-dependent evolution of the electric potential (in reference to the outer ring) at each point. 

\begin{figure}[h]
	\centering
	\includegraphics[width=1\linewidth]{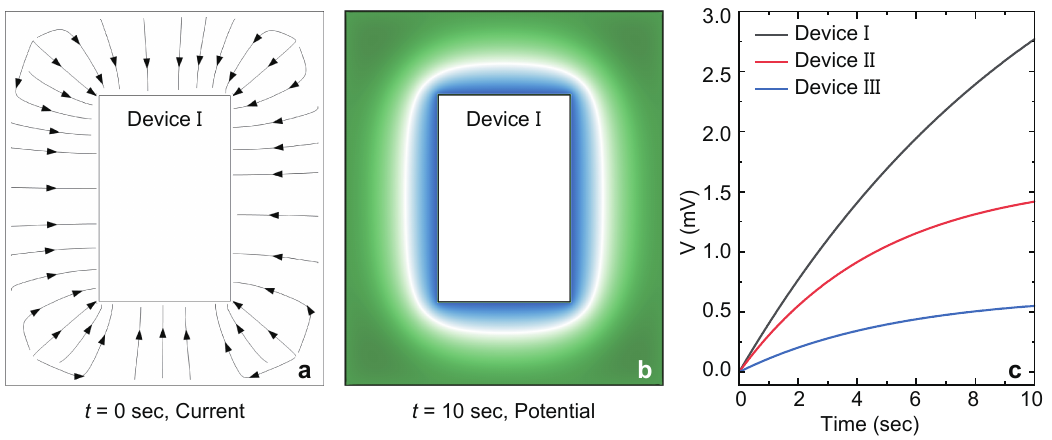}
	\caption{The typical distribution of simulated current (a) and electric potential (b). (c) Simulated $V$ as a function of time for the three device sizes listed in Tab. \ref{Tab_5_SampleList}. 
	}
	\label{fig_5_3Simulation_AreaFactor}
\end{figure}

In our numerical calculation, we use $\mathrm{d}B/\mathrm{d}t = 0.1$ T/min, $\sigma_\mathrm{xx} = 1\times10^{-11}\,\mathrm{\Omega}^{-1}$ and $\sigma_\mathrm{xy}=e^2/h$ as the quantum conductance value at the IQH state $\nu=1$. The external capacitance is set to $C = 1$ nF. The typical current and potential distribution is shown in Fig.~\ref{fig_5_3Simulation_AreaFactor}(a) and (b). We summarize the time evolution of the voltage $V$ between the inner and outer contacts in Fig.~\ref{fig_5_3Simulation_AreaFactor}(c). We can calculate the pumping current using Eq.~\eqref{eq_5_Cal_I_CV} and then deduce the effective area according to Eq.~\eqref{eq_5_Cal_Ip_solution}. Furthermore, we can perform an exponential fit to the time-dependent voltage variation $V(t) \propto 1-\exp(-t/\tau)$ to calculate the time constant $\tau$. Similar to Eq.~\eqref{eq_5_Cal_V_solution}, the geometry factor $\beta$ can be derived through Eq.~\eqref{eq_5_Cal_V_solve} as $\beta=\frac{C}{\tau\sigma_\mathrm{xx}}$. Our numerical simulations were conducted for all devices and the effective area $S$ and geometry factor $\beta$ is summarized in Tab.~\ref{Tab_5_SampleList}.

\section{Mitigation of experimental imperfections}

There are several mechanisms that can drive electrons through the contacts as we sweep the magnetic field through quantum Hall plateaus. Fortunately, most of the apparatus imperfections cause a current that is independent of the polarity of $\mathrm{d}B/\mathrm{d}t$. In our experiments, these effects can be eliminated by averaging results taken from the up- and down-sweeps. 

\subsection{Residual current and SAW drag current} 
\label{sec_Residual}

First of all, in addition to the pumping current $I_\mathrm{p}$ (and the accompanying reverse leakage current $I_\mathrm{leak}$), the measurement cables and instruments also exhibit an intrinsic residual current $I_\mathrm{res}$. This residual current also causes a charge accumulation. The accumulation rate ($I_\mathrm{res}$) is independent of both the magnetic field sweep rate and the sample's condition but is rather determined by the instrument status during measurement. 

\begin{figure}[!htbp]
	\includegraphics[width=1\textwidth]{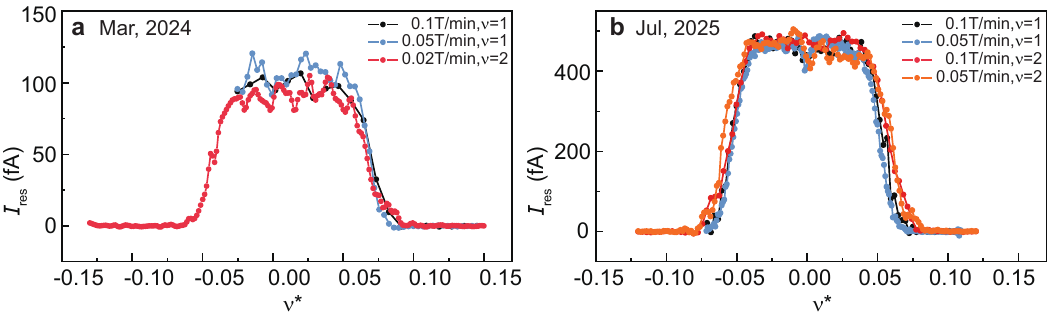}
	\caption{\label{fig_5_4Result_2_Residual} 
            Residual current $I_\mathrm{res}$ from two different measurement cycles of the Device \uppercase\expandafter{\romannumeral 1} that are separated by an interval of more than one year. $I_\mathrm{res}$ is identical for both $\nu=1$ (black, blue curves) and $\nu=2$ (red, orange curves). It is also independent of the magnetic field sweep rate.
		}
\end{figure}

The residual signal in Fig. 2 (a) is likely caused by the finite and drifting zero-input offset of the electrometer (the input voltage bias when the output equals zero). It may introduce a voltage bias of $\sim100$ \textmu V at the capacitor and slowly charges the contacts. When magnetic flux density increases, $I_\mathrm{res}$ flows along the same direction as the pumping current. When magnetic flux density decreases, the pumping current reverses its direction but $I_\mathrm{res}$ remains unchanged. As a result, the current measured during the up- and down-sweep directions in the original data is not the same. We can take the antisymmetric component with respect to the sweep direction as the pumping signal, and the symmetric component as the residual current signal.

$I_\mathrm{res}$ does not change during the measurements period of several days. Figure \ref{fig_5_4Result_2_Residual} shows the residual current of the same sample during two cool-down cycles separated by more than one year. It is clearly observed that $I_\mathrm{res}$ remains constant within each measurement period of several days, independent of both the magnetic field sweep rate and the filling factor. However, it changes when the status of the fridge and instruments changes during different cool-down cycles and long time scales.

\begin{figure}[!htbp]
	\centering
	\includegraphics[width=1\linewidth]{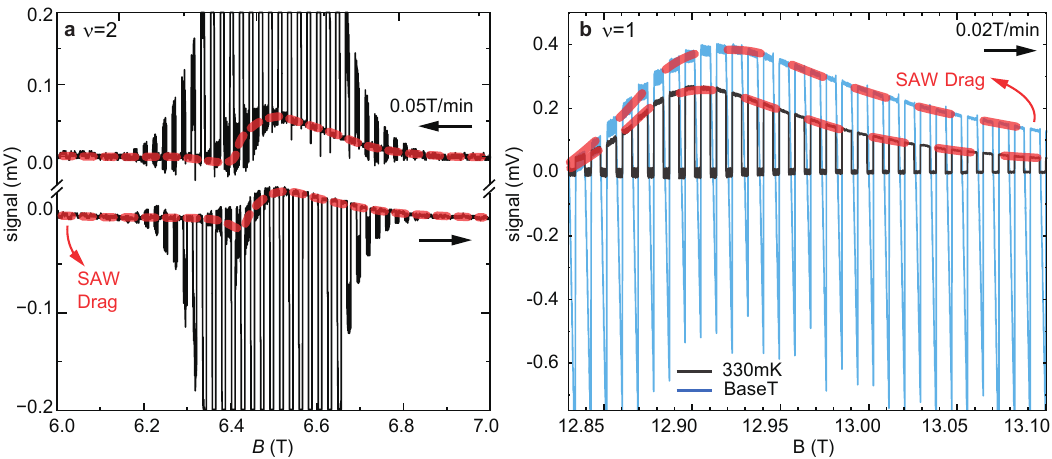}
	\caption{
		The measured signals at (a) $\nu=2$ and (b) $\nu=1$ IQH. The experiment was conducted in Device \uppercase\expandafter{\romannumeral 1}. (a) The drag signals (highlighted by the orange dashed curve) are independent of the magnetic field sweeping direction and significantly smaller than the pump signals. (b) When the temperature increases, the pump signal almost disappears (signal is nearly zero when SAW is off), while the drag signal only slightly decreases.
	}
	\label{fig_5_2Method_4_SAWDrag}
\end{figure}

A recent study reports phonon-drag effect even when the quantum Hall effect is strong\cite{WangRenfei2025}, so that the SAW used in this study may leave a residual voltage across the sample. The drag signal observed in this study is shown in Fig.~\ref{fig_5_2Method_4_SAWDrag}. Both of these effects are independent of the magnetic field sweep direction.

\begin{figure}[!htbp]
	\includegraphics[width=1\textwidth]{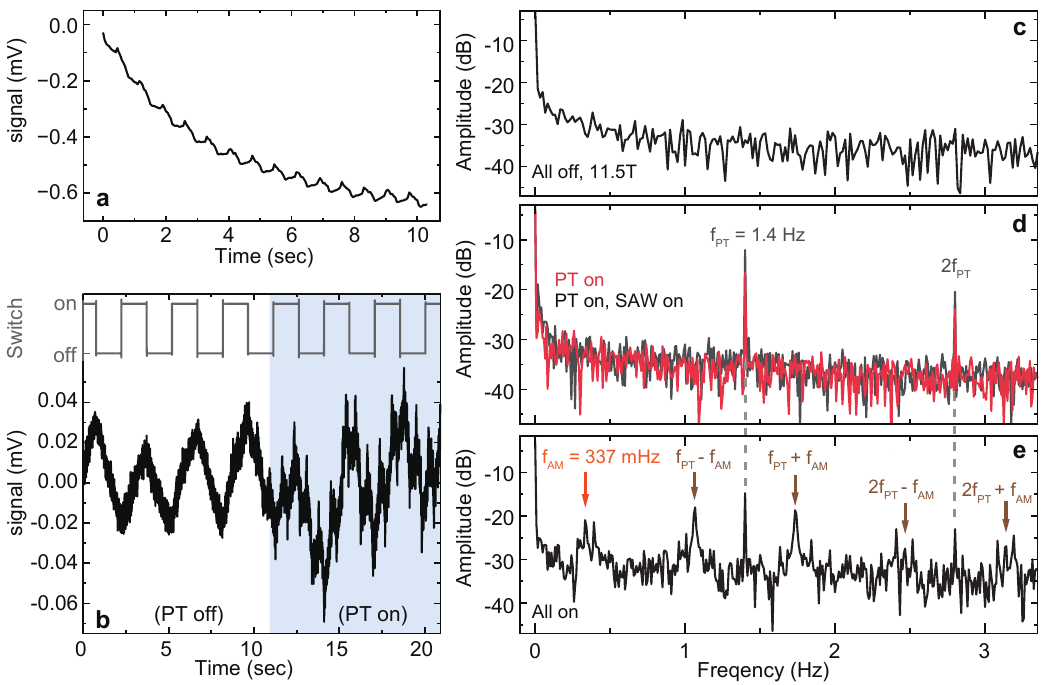}
	\caption{\label{PTandnoise} 
        (a) Typical charging signal. The oscillation is caused by the pulse tube (PT). (b) The oscillation disappears when the PT is turned off, and clean charging and discharging cycles can be seen. (c) The noise spectrum when both the SAW and PT are turned off. (d) The noise spectrum with PT when the SAW is either on or off. (e) The noise spectrum of the typical measured signal.
		}
\end{figure}

\subsection{Other non-ideal effects}

Furthermore, the wire inside the fridge inevitably encloses a finite area, so that the finite $\mathrm{d}B/\mathrm{d}t$ generates a potential between the two contacts through Faraday effects. This potential is a problem in AC measurements. Fortunately, it is not an issue in our experiments. The pumping process is related to the time-gradient of the capacitance voltage, $\mathrm{d}V/\mathrm{d}t$, while the Faraday effect only causes a DC bias in it. The non-zero leaking leads to a similar behavior to the pumping, but its amplitude is orders of magnitude smaller. For example, the pumping current is about 500 fA at $\nu=1$ when $dB/dt=1.7$ mT/s. Meanwhile, the $\sim$ 1 mV Farady voltage can only generate a $\sim$ 10 fA leaking current through the insulating corbino ring, whose conductance is $\lesssim 10^{-12}$ S near integer filling factors. While the longitudinal conductance changes by orders of magnitude near integer fillings, we observe a flat plateau in the measured pumping coefficient, further confirming that the Faraday effect is not an issue in our experiment.

The 2DES density varies when its chemical potential changes during sweeping the magnetic field, if the sample has a proximate gate which draws (pushes) electrons into (out of) the 2DES. The 2DES has a discontinuity at exact integer filling factors where the electron chemical potential jumps between discrete Landau levels. Inserting (removing) magnetic flux through the quantum Hall liquid generates a dissipationless eddy current around the sample $I\propto d\phi/dt$. The 2DES chemical potential varies by $\rho_{xy} I$ according to the sweep direction. The imperfection of the 2DES samples such as the Schottky junction formed between the metallic contacts and the quantum Hall bulk, the non-uniform electron density, etc., usually lead to the redistribution of the electrons. Our sample has no proximate gate which translates the chemical potential into electron density. By switching the two current probe leads, we can measure the amount of electrons passing through the inner and outer contacts. Their values are equal so that the 2DES density remains constant.

\subsection{Detection resolution}

The detection resolution is limited by the noise of the charge sensor. The white noise level is sufficiently low so that we can easily resolve $\nu$V-level signal, corresponding to $\sim$ 1000 electrons drained to the capacitor; see Fig.~\ref{PTandnoise}. However, one of the two capacitor leads is essentially floating when the sample exhibits strong quantum Hall effect. Therefore, the measured signal is highly fragile from external perturbations. The major disturbance is the vibration of the dilution refrigerator which is seen as a 1.4 Hz oscillation in addition to the slowly charging curve in Fig.~\ref{PTandnoise} a. This vibration is caused by the pulse tube. In Fig.~\ref{PTandnoise} b, it disappears when we temporally turn off the pulse tube and clean charging and discharging signals can be seen. This is a dangerous operation as the superconducting magnet temperature rises and will quench in one or two minutes. The noisy oscillation reappears as soon as the pulse tube is turned on again. The amplitude of this oscillation is smaller when the SAW is on and the sample conductivity increases, resulting in a complicated spectrum; see Fig.~\ref{PTandnoise} c and d. Therefore, this disturbance is unavoidable and limits our detection limit of the pumping rate to about $10^4$ electron per second.

\section{Definition of $\nu_{\rm CF}$, $\ell_{\rm CF}$ and $a_\mathrm{CF}^*$} 

On quantum Hall plateaus, we can define the deviation of filling factor $\nu^{*}=\nu-\nu_{0}$ with $\nu_{0}=\sigma_{xy}/(e^{2}/h)$. For integer plateaus, the areal density of excitations is $n^{*}=\frac{\nu^{*}}{\nu} n$ with $n$ being the electron density. If each excitation occupies a circle with radius $r$, the average distance between two excitations is $a^{*}=2r=\sqrt{4/(\pi n^{*})}$. 

Since the electron densities in our samples are large, the fractional QHSs at $\nu=4/3$ and $5/3$ should be spin-polarized: the spin-up $N=0$ LL is fully populated and the spin-down $N=0$ LL is partially filled. The CF filling factor $\nu_{\rm CF}$ is related to the electron filling factor via $\nu-1=\frac{\nu_{\rm CF}}{2\nu_{\rm CF}\pm 1}$. For the $\nu=4/3$ and $5/3$ states, we have $\nu_{\rm CF}=1$ with positive sign and $\nu_{\rm CF}=2$ with negative sign, respectively. The effective magnetic field for the CFs in the vicinity of $\nu=4/3$ is $B_{\rm eff} = B/(1+2\nu_{\rm CF})$ and that in the vicinity of $\nu=5/3$ is $B_{\rm eff} = B/(1-2\nu_{\rm CF})$. We can thus define the CF magnetic length $\ell_{\rm CF}=\sqrt{\hbar/(eB_{\rm eff})}$. As the electron filling factor deviates from $4/3$ or $5/3$, the CF filling factor also deviates from $1$ or $2$. We thus define $\nu^{*}_{\rm CF}=\nu_{\rm CF}-1$ and $\nu^{*}_{\rm CF}=\nu_{\rm CF}-2$ for the two states. The CF areal density $n_{\rm CF}$ is determined by the electron density in the spin-down $N=0$ Landau level as $n_{\rm CF}= n(\nu-1)/\nu$. When the CF filling factor deviates from integers, the density of excess CFs is $n^{*}_{\rm CF}= n_{\rm CF} \nu^{*}_{\rm CF}/\nu_{\rm CF}$. This tells us that the average distance between excess CFs is $a^{*}_{\rm CF} = \sqrt{4/(\pi n^{*}_{\rm CF})}$. 


\section{Additional pumping data}

\begin{figure}[!htbp]
	\includegraphics[width=0.5\textwidth]{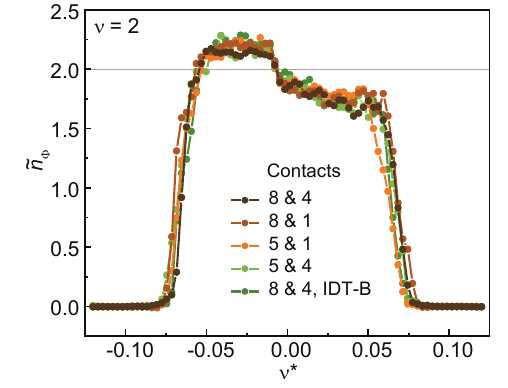}
	\caption{\label{fig_5_4Result_x_DiffContactDiffIDT} 
            The $\tilde{n}_{\rm \Phi}$ results is measured using different contact pairs and IDT. All other contact electrodes are floating during the measurement. The numbering of the contacts and IDTs follows Fig.~\ref{figS1}. The experiment is conducted in the $\nu=2$ plateau of the Device \uppercase\expandafter{\romannumeral 1}.
		}
\end{figure}

Figure~\ref{fig_5_4Result_x_DiffContactDiffIDT} presents $\tilde{n}_{\rm \Phi}$ measured at $\nu=2$ using IDT-A and different contact pairs. As long as the measurement is conducted between the inner and outer contacts, there is no difference in the results regardless of the contact selection. Through this result, we confirm that the signal we measured originates from radial charge pumping between the inner and outer contacts, and there is no signal along the angular direction between the contacts of the inner ring (or outer ring). We also perform experiments when SAW is excited by the IDT-B. The results indicate that the role of SAW is merely releasing the charges accumulated between the capacitor plates, and the result is independent of the SAW propagation direction.

\begin{figure}[!htbp]
	\includegraphics[width=\textwidth]{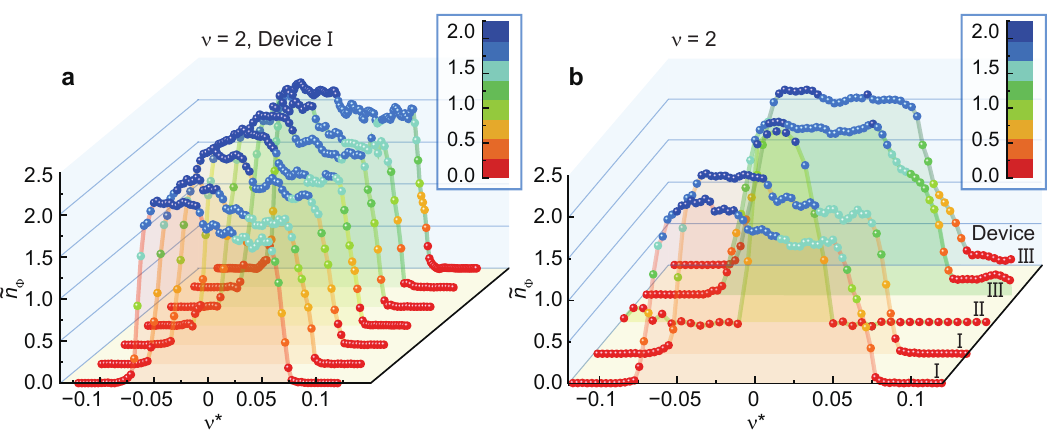}
	\caption{\label{fig_5_4Result_5_DiffSize} 
            Charge pumping results at filling factor $\nu=2$. (a) Data from Device \uppercase\expandafter{\romannumeral 1} using different sweeping rates.
            (b) Results from different devices.
		}
\end{figure}

Figure \ref{fig_5_4Result_5_DiffSize} (a) shows results for $\nu=2$ in Device \uppercase\expandafter{\romannumeral 1}. Different curves correspond to different measurement conditions: the magnetic field sweeping rates range from 0.05 to 0.1 T/min, the SAW input power is either -35 or -40 dBm, and the SAW switching frequency varies from 137 to 177 mHz. The measurement results of different devices at the $\nu=2$ plateau are shown in Fig.~\ref{fig_5_4Result_5_DiffSize} (b). These three samples with significantly different sizes exhibit identical plateau height. The difference in plateau width is likely because of the different quality. For Device \uppercase\expandafter{\romannumeral 3}, the ``tail'' observed on the high-filling-factor side stems from a low-density contact of the sample. This phenomenon is also shown in the standard transport measurements of this sample.

\section{Additional $\sigma_\mathrm{xx}$  data}

\begin{figure}[!htbp]
	\includegraphics[width=1\textwidth]{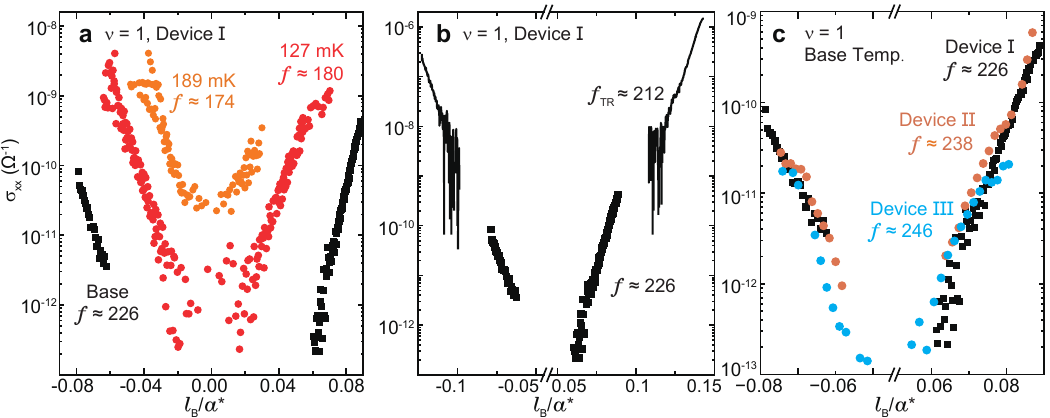}
	\caption{\label{Fig_S8_Sigma_Temp} 
            (a) $\sigma_\mathrm{xx}$ as a function of $\ell_{\rm B}/a^*$ at different temperature. The parameter $f$ only varies from 226 to 180 while $\sigma_\mathrm{xx}$ increases by 3 to 4 orders of magnitude. (b) The $\sigma_\mathrm{xx}$ obtained by our method and conventional transport method follows similar exponential dependence with nearly the same $f$. (c) $\sigma_\mathrm{xx}$ taken from different devices at base temperature.
		}
\end{figure}

Figure \ref{Fig_S8_Sigma_Temp}a shows the relation between $\sigma_\mathrm{xx}$ and $\ell_{\rm B}/a^*$ near $\nu=1$ at different temperatures. When temperature increases, $\sigma_\mathrm{xx}$ increases while its dependence on $\ell_{\rm B}/a^*$ remains nearly unchanged. $\sigma_\mathrm{xx}$ near $\nu^*=0$ at low temperatures are too small to be measured directly. We can determine its value by fitting the data in Figure \ref{Fig_S8_Sigma_Temp}a and extrapolating to $\ell_{\rm B}/a^*=0$. The results are included as open symbols in Fig. 3b. Figure \ref{Fig_S8_Sigma_Temp}b compares $\sigma_\mathrm{xx}$ measured by our method and conventional quasi-DC transport measurement. The results have the same dependence and nearly the same slope $f\approx 226$ and 212. Fig.~\ref{Fig_S8_Sigma_Temp}c shows that
$\sigma_\mathrm{xx}$ taken from different devices are consistently the same, indicating that this dependence is universal. 

\begin{figure}[!htbp]
	\includegraphics[width=\textwidth]{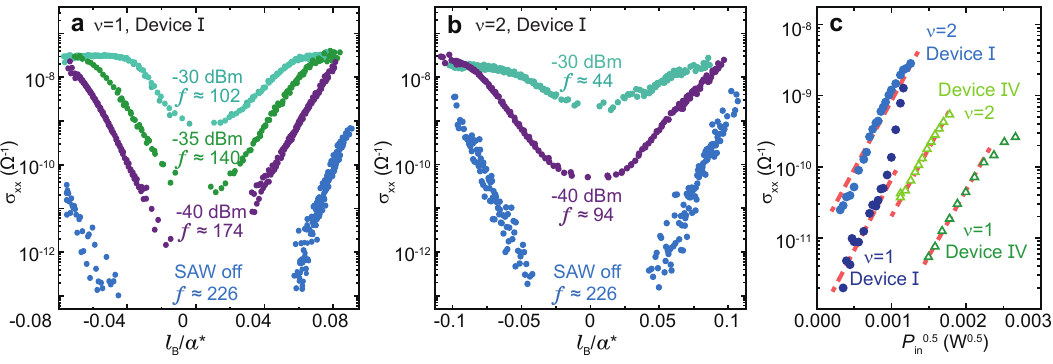}
	\caption{\label{fig_5_5Sigma_4_SAWon} 
            (a,b) The conductivity $\sigma_\mathrm{xx}$ vs. $\ell_{\rm B}/a^*$ at different SAW powers near (a) $\nu=1$ and (b) $\nu=2$. The $P_\mathrm{in}$ is the input RF power to excite SAW. (c) The $\sigma_\mathrm{xx}$ vs. $\sqrt{P_\mathrm{in}}$ dependence at exact integer filling $\nu=1$ and 2. $\sqrt{P_\mathrm{in}}$ is proportional to the SAW's piezoelectric field $V_{\rm SAW}$. All experiments were conducted at the base temperature of the dilution fridge (approximately 50 mK). 
		}
\end{figure}

The conductivity $\sigma_\mathrm{xx}$ of the quantum Hall insulator increases when the SAW is turned on. Our SEA scheme utilizes this phenomenon to release the accumulated charges and reset the voltage back to zero. The $\sigma_\mathrm{xx}$ in the presence of SAW can also be derived from the corresponding time constant of this discharging process, and the measured results are shown in Fig.~\ref{fig_5_5Sigma_4_SAWon}. Figure~\ref{fig_5_5Sigma_4_SAWon}(a) and (b) present the results under different SAW excitation powers $P_\mathrm{in}$ for the Device \uppercase\expandafter{\romannumeral 1} at $\nu=1$ and $2$, respectively. Surprisingly, in addition to the increasing of $\sigma_\mathrm{xx}$, the parameter $f$ which is nearly independent of the filling factor, sample quality and temperature, decreases by a factor of 2 when the SAW power increases. This indicates that the SAW-induced conductivity has a non-trivial origin. 

In Fig.~\ref{fig_5_5Sigma_4_SAWon}(c) we investigate the $\sigma_\mathrm{xx}$ as a function of the input SAW power $P_{\rm in}$. Note that the $x$-axis is $\sqrt{P_{\rm in}}$, which is proportional to the piezoelectric field $V_{\rm SAW}$ associated with the SAW. \footnote{The input radio frequency power $P_\mathrm{in}$ and the SAW power $P_\mathrm{SAW}$ are different. This difference arises from cable attenuation, power dissipation caused by the 50 $\Omega$ resistor connected with the IDT, and the electro-acoustic conversion efficiency during SAW excitation. The insertion loss between the transmitting and receiving IDTs is -34 dB.} These experimental data suggest that $\sigma_\mathrm{xx}$ exhibits an exponential dependence on $V_{\rm SAW}$. The devices I and IV have the same geometry and IDT structure, so that we expect the same $V_{\rm SAW}$-$P_{\rm in}$ conversion. In Fig.~\ref{fig_5_5Sigma_4_SAWon}(c), it is surprising that the data taken at different filling factors and from devices with different quality have similar slope. As far as we know, there is no proper explanation for experiments in this regime yet.



\end{document}